\documentclass{PoS}
\usepackage{axodraw}
\usepackage{slashed}
\title{Quantum field theory in four dimensions}

\ShortTitle{FDR}

\author{\speaker{Roberto Pittau}
\thanks{Work performed in the framework of the ERC grant 291377, ``LHCtheory: Theoretical predictions and analyses of LHC physics: advancing the precision frontier'' and of the MICINN project FPA2011-22398 (LHC@NLO).}\\
        Departamento de F\'isica Te\'orica y del Cosmos and CAFPE,\\
        Campus Fuentenueva s.n., Universidad de Granada, E-18071 Granada, Spain\\
        E-mail: \email{pittau@ugr.es}}

\abstract{I review the current status of FDR, the recently introduced Four-Dimensional Regularization/Renormalization approach to ultraviolet divergences in Quantum Field Theory. FDR also regulates infrared and collinear infinities in the intermediate steps of the calculation.}

\FullConference{Proceedings of the Corfu Summer Institute 2012 "School and Workshops on Elementary Particle Physics and Gravity"\\
		September 8-27, 2012\\
		Corfu, Greece}

\begin{document}
\def\d#1{D_{#1}}
\def\db#1{\bar D_{#1}}
\def\tld#1{\tilde {#1}}
\newcommand{\qbarslash}{\,\overline{\slashed{q}}}
\newcommand{\fdr}{\,[\de^4q]}
\newcommand{\I}{\,\mathfrak{I}}
\newcommand{\D}{\,{D}}
\newcommand{\Dbar}{\,\overline{{D}}}
\newcommand{\qbar}{\,\overline{q}}
\newcommand{\amp}{\,\mathcal{M}}
\newcommand{\mur}{\mu_{\scriptscriptstyle R}}
\newcommand{\nl}{\nonumber \\}
\newcommand{\bfig}{\begin{center}\begin{picture}}
\newcommand{\efig}[1]{\end{picture}\\{\small #1}\end{center}}
\newcommand{\flin}[2]{\ArrowLine(#1)(#2)}
\newcommand{\ghlin}[2]{\DashArrowLine(#1)(#2){5}}
\newcommand{\wlin}[2]{\DashLine(#1)(#2){2.5}}
\newcommand{\zlin}[2]{\DashLine(#1)(#2){5}}
\newcommand{\glin}[3]{\Photon(#1)(#2){2}{#3}}
\newcommand{\gluon}[3]{\Gluon(#1)(#2){5}{#3}}
\newcommand{\lin}[2]{\Line(#1)(#2)}
\newcommand{\sof}{\SetOffset}
\newcommand{\bqa}{\begin{eqnarray}}
\newcommand{\eqa}{\end{eqnarray}}
\section{Introduction}
 The computation of Radiative Corrections (RCs) is very demanding from a technical point of view. A lot of work has been recently devoted to deal, in an efficient way, with 1-loop processes at large multiplicities~\cite{Ossola:2006us,Berger:2008sj,Giele:2008ve,Ellis:2011cr}, and progress has been achieved also in the field of the multi-loop calculations~\cite{Mastrolia:2012an,Mastrolia:2011pr,Badger:2012dv,Johansson:2012zv,Kleiss:2012yv}. Most of the difficulties are triggered by the usual treatment, in the framework of Dimensional Regularization (DR)~\cite{'tHooft:1972fi}, of the infinities arising in the intermediate steps of the calculation, so that several attempts have been tried out to find four-dimensional alternatives to the DR treatment of the UV infinities, such as differential renormalization~\cite{Freedman:1991tk}, constrained differential renormalization~\cite{delAguila:1997kw,delAguila:1998nd}, which both work in the coordinate space, implicit renormalization~\cite{Battistel:1998sz,Cherchiglia:2010yd}, symmetry preserving regularization~\cite{Cynolter:2010ei} and LR~\cite{Wu:2003dd}, directly applicable in the momentum space.

In a recent work, the FDR approach~\cite{Pittau:2012zd} has been proposed in which the UV problem is solved by simply re-interpreting the loop integrals appearing in the calculation. They are {\em defined} in such a way that infinities do not occur. The price to pay is the appearance of an arbitrary scale, called  $\mu$, which plays the role of the renormalization scale. Technically speaking, infinities never appear, and the procedure works because the FDR re-interpretation respects, by construction, shift and gauge invariance. In this contribution, I review the state-of-the-art of the FDR approach.
\section{The FDR integral}
The UV convergence of a loop integral can be improved by a repeated use of the  
\emph{partial fraction identity}
\bqa
\label{partial_fractional_identities}
	\frac{1}{\Dbar_i}  = \frac{1}{\qbar^2}
		\Bigg(1+\frac{d_i}{\Dbar_i} \Bigg)	\,,
\eqa
with
\bqa
\label{defd}
D_i ~=~ q^2-d_i\,,~~d_i ~=~ M^2_i-p^2_i- 2 (q\cdot p_i)\,,
\eqa
and
\bqa
\Dbar_i~=~\D_i-\mu^2\,,~~\qbar^2~=~q^2-\mu^2\,,
\eqa
where $q$ is the loop momentum and $\mu$ an arbitrary, vanishing, scale needed to avoid the appearance of possible infrared divergences in the second term of the r.h.s of eq.~(\ref{partial_fractional_identities}). 
Consider, as an example, the 1-loop integral~\footnote{I take $p_0= 0$.} 
\bqa
\label{eq:logdivint}
	\I =\int d^4q \frac{1}{\D_0 \D_1}\,.
\eqa
After promoting
\bqa
\label{shift}
 q^2 \to \qbar^2\,, 
\eqa
namely
\bqa
D_i  \to \Dbar_i\,,
\eqa
eq.~(\ref{partial_fractional_identities}) yields
\bqa
\label{eq:eq1}
\I
	= \lim_{\mu\rightarrow0}\; 
             \int d^4q \,\Bigg( 
			\Bigg[\frac{1}{\qbar^4}\Bigg]
			+\frac{d_1}{\qbar^4\Dbar_1}
			+\frac{d_0}{\qbar^2\Dbar_0\Dbar_1} 
			\Bigg)\,.
\eqa
Only the term between square brackets in eq.~(\ref{eq:eq1}) leads to UV divergences. On the other hand, any physically relevant scale is contained in the remaining part. One can therefore {\em define} the FDR integral by simply dropping the divergent integrand:
\bqa
\I^{\rm{FDR}} = \int [d^4q] 
\frac{1}{\Dbar_0 \Dbar_1} \equiv
	\lim_{\mu\rightarrow0}\; \int d^4 q \,\Bigg(\; 
			 \frac{d_1}{\qbar^4\Dbar_1}
			+\frac{d_0}{\qbar^2\Dbar_0\Dbar_1} 
			\;\Bigg)\ \Bigg|_{\mu=\mu_R} .
\eqa
The final identification $\mu= \mu_R$ effectively eliminates the dependence on the original cut-off, as explained in detail in~\cite{Pittau:2012zd}.
The extension to more loops is straightforward. Any multi-loop integrand
$J = J_V + J_F$ can be always be split into terms which only contains $\mu$, called vacuum configurations $J_V$, and a finite part $J_F$, resulting in
\bqa
       \label{eq_FDR}
	\I^{\rm{FDR}}_{\ell} 
	&=&
	{\displaystyle \int \prod_{i=1}^\ell [ d^4 q_i]}\;
	J(\{\qbar^2\}) 
	~\equiv~ \lim_{\mu\rightarrow0} \,
	{\displaystyle \int \prod_{i=1}^\ell d^4 q_i}\;
	J_F( \{\qbar^2 \}) 
	\Bigg|_{\mu=\mu_R}
	. 
\eqa
When computing Feynman diagrams, it is important to realize that the shift
in eq.~(\ref{shift}) should be performed in both denominators {\em and} numerators; and that $\mu^2$ integrals such as 
\bqa
\label{intmu}
\I^{\rm{FDR}}(\mu^2) = 
	\int[d^4q]\frac{\mu^2}{\Dbar_0\Dbar_1\Dbar_2}\,,
\eqa
require the {\em same } denominator expansion needed to subtract the vacuum configurations from
\bqa
\int[d^4q]\frac{q^2}{\Dbar_0\Dbar_1\Dbar_2}\,.
\eqa
That ensures, for example, identities between integrals, such as~\cite{Donati:2013iya}
\bqa
	\int[d^4q] \frac{\qbar^2}{\Dbar_0\Dbar_1\Dbar_2} = 
	\int[d^4q]\frac{1}{\Dbar_1\Dbar_2} + 
        \int[d^4q]\frac{d_0}{\Dbar_0\Dbar_1\Dbar_2}\,,
\eqa
which are essential to keep the cancellations needed to prove the Ward Identities in gauge theories.

\section{Shift invariance of the FDR integral}
The definition in eq.~(\ref{eq_FDR}) implies invariance under any change of variable, as it becomes evident by considering the FDR integral as a difference between an integral $\I^{\rm{DR}}_{\ell}$, regulated (for example~\footnote{One is free to choose any regulator.}) in dimensional regularization, and its vacuum configurations:
\bqa
 \label{FDR_as_a_difference}
	\I^{\rm{FDR}}_{\ell} = \I^{\rm{DR}}_{\ell} -
	\lim_{\mu\rightarrow0} \,
	\mu_R^{-\ell\epsilon}\int \prod_{i=1}^\ell d^n q_i\;
	J_V (\{ \qbar^2\}) 
	\Bigg|_{\mu=\mu_R}.
\eqa
Shift invariance can be easily verified explicitly. 
Consider, e.g.
\begin{eqnarray}
I_\alpha       ~=~ \int[d^4q] \frac{q_\alpha}{\bar D_p^2}\,,~~~
I^\prime_\alpha ~=~ \int[d^4q] \frac{(q_\alpha-p_\alpha)}{\bar D^2}\,, 
\end{eqnarray}
with
\begin{eqnarray}
\bar D_p ~=~ (q+p)^2-M^2-\mu^2\,,~~~ \bar D   ~=~ q^2-M^2-\mu^2\,.
\end{eqnarray}
One must have $I_\alpha= I^\prime_\alpha$, which can be proved either directly, from the FDR definition of $I_\alpha$ and $I^\prime_\alpha$, or indirectly, by subtracting vacuum configurations from the corresponding dimensionally regulated integrals.

The direct computation of $I_\alpha$ requires the following expansion of its
integrand
\begin{eqnarray}
\label{eq:eq2}
\frac{q_\alpha}{\bar D_p^2} &=& 
 \left[\frac{q_\alpha}{\bar q^4} \right]
-4\left[\frac{(q\cdot p)q_\alpha}{\bar q^6} \right] \nonumber \\
&+& d_0 q_\alpha 
\left(\frac{1}{\bar q^4 \bar D_p}+ \frac{1}{\bar q^2 \bar D_p^2} \right)
-2 q_\alpha(q \cdot p) d(q) 
\left(\frac{2}{\bar q^6 \bar D_p}+ \frac{1}{\bar q^4 \bar D_p^2} \right)\,,
\end{eqnarray} 
with
\bqa
\bar q^2 = q^2-\mu^2\,,~~~~ 
   d(q) = d_0 -2(q\cdot p)\,,~~~~ 
   d_0  = M^2-p^2\,,
\eqa
and where the terms between square brackets are divergent.
Therefore
\bqa
I_\alpha = d_0 (J_{1\alpha}-2J_{2\alpha}) + 4 J_{3\alpha}\,,
\eqa
with
\begin{eqnarray}
J_{1\alpha} &=& \lim_{\mu \to 0} \int d^4q\, q_\alpha \left(\frac{1}{\bar q^4 \bar D_p}+ \frac{1}{\bar q^2 \bar D_p^2} \right)\,, \nonumber \\
J_{2\alpha} &=& \lim_{\mu \to 0} \int d^4q\, q_\alpha (q \cdot p)
\left(\frac{2}{\bar q^6 \bar D_p}+ \frac{1}{\bar q^4 \bar D_p^2} \right)
\,, \nonumber \\
J_{3\alpha} &=& \lim_{\mu \to 0} \int d^4q\, q_\alpha (q \cdot p)^2
\left(\frac{2}{\bar q^6 \bar D_p}+ \frac{1}{\bar q^4 \bar D_p^2} \right)\,.
\nonumber
\end{eqnarray}
Computing the previous integrals gives
\begin{eqnarray}
I_\alpha= i \pi^2 p_\alpha\,\ln \frac{M^2}{\mu^2}\,. \nonumber
\end{eqnarray}

The starting point for the indirect computation of $I_\alpha$ is instead
\begin{eqnarray}
\label{eq2}
I_\alpha= \lim_{\mu \to 0} \int d^nq \,q_\alpha
\left\{  
  \frac{1}{((q+p)^2-M^2-\mu^2)^2} 
- \left[\frac{1}{(q^2-\mu^2)^2} 
- 4\frac{(q\cdot p)}{(q^2-\mu^2)^3} \right]
\right\}\,,
\end{eqnarray}
namely the l.h.s. of eq.~(\ref{eq:eq2}) subtracted by the divergent integrands
appearing in the r.h.s. An easy calculation gives
\begin{eqnarray}
I_\alpha= i \pi^2 p_\alpha\,\ln \frac{M^2}{\mu^2}\,. \nonumber
\end{eqnarray}
Analogously, both direct and indirect computations of $I^\prime_\alpha$ confirm 
that
\begin{eqnarray}
I^\prime_\alpha= i \pi^2 p_\alpha\,\ln \frac{M^2}{\mu^2} = I_\alpha \,. \nonumber
\end{eqnarray}
\section{Fermions in FDR}
In the presence of strings of Dirac matrices, the replacement of eq.~(\ref{shift}) in the numerator of the amplitude is equivalent to a shift
\bqa
\slashed{q}\rightarrow \qbarslash \equiv q \pm \mu\,,
\eqa
directly performed in the fermionic string~\cite{Donati:2013iya}, where $\qbarslash$ is defined according to its position:
\bqa \label{eq_fermion_prescription}
	( \ldots 
		\, \qbarslash \; \gamma^{\alpha_1}\ldots\gamma^{\alpha_n}
		\qbarslash\, \ldots )
	= 
	(\dots 
		\, (\slashed{q}+\mu) \; \gamma^{\alpha_1}\ldots\gamma^{\alpha_n} (\slashed{q}-(-)^{n}\mu)  \ldots ) \,.
\eqa
To prove the equivalence, one should also make use the fact the FDR integrals involving odd powers of $\mu$ in the numerator vanish~\cite{Pittau:2012zd}.

If chirality matrices are involved, a gauge invariant treatment~\cite{Jegerlehner:2000dz} requires their anticommutation at the beginning (or the end) of open strings before replacing $\slashed{q}\rightarrow \qbarslash$. In the case of closed loops, $\gamma_5$ should be put next to the vertex corresponding to a potential non-conserved current. This reproduces the correct coefficient of the triangular anomaly, as observed in~\cite{Pittau:2012zd}. 

\section{FDR at two-loop}
As an example of two-loop FDR regularization, consider the integral 
\begin{equation}
\int [d^4q_1][d^4q_2] \frac{1}{\bar D_1\bar D_2\bar D_{12}}\,,
\end{equation}
where the propagators are given by
\begin{eqnarray}
\label{eq:2loop}
\bar D_1   &=& \bar{q}_1^2-m_1^2\,,\nonumber \\ 
\bar D_2   &=& \bar{q}_2^2-m_2^2\,,\nonumber \\
\bar D_{12} &=& \bar{q}_{12}^2-m_{12}^2\,.
\end{eqnarray}
In the same spirit of the one-loop case, divergent integrands 
can be subtracted before integration by means of eq.~(\ref{partial_fractional_identities}), or 
\bqa
\frac{1}{\bar{q}_{12}^2} &=& \frac{1}{\bar{q}_2^2}
             -\frac{q_1^2+2(q_1 \cdot q_2)} {\bar{q}_2^2\bar{q}_{12}^2}\,,
\eqa
resulting in the following expression:
\begin{eqnarray}
\label{eq:eq22}
\int [d^4q_1][d^4q_2] \frac{1}{\bar D_1\bar D_2\bar D_{12}} &=& \nl
\lim_{\mu \to 0} 
\int d^4q_1 \int d^4q_2
&&\!\!\!\!\!\!\!\!\!\!\!\left(
 \frac{m_1^2 m_2^2}{(\bar D_1 \bar{q}_1^2)(\bar D_2 \bar{q}_2^2)\bar{q}_{12}^2} 
+\frac{m_1^2 m_{12}^2}{(\bar D_1 \bar{q}_1^2)\bar{q}_{2}^2(\bar D_{12} \bar{q}_{12}^2)} 
+\frac{m_2^2 m_{12}^2}{\bar{q}_{1}^2(\bar D_2 \bar{q}_2^2)(\bar D_{12} \bar{q}_{12}^2)} 
\right.
\nonumber \\
&-& 
  m_1^4 \frac{q_1^2+2(q_1 \cdot q_2)}{(\bar D_1 \bar{q}_1^4)\bar{q}_2^4 \bar{q}_{12}^2}
- m_2^4 \frac{q_2^2+2(q_1 \cdot q_2)}{\bar{q}_1^4(\bar D_2 \bar{q}_2^4) \bar{q}_{12}^2} 
- m_{12}^4 \frac{q_{12}^2-2(q_1 \cdot q_{12})}{\bar{q}_1^4 \bar{q}_{2}^2
(\bar D_{12} \bar{q}_{12}^4)}
\nonumber \\
&+&
\left. \left.
\frac{m_1^2 m_2^2 m_{12}^2}{(\bar D_1 \bar{q}_1^2)(\bar D_2 \bar{q}_2^2)(\bar D_{12} \bar{q}_{12}^2)}
\right)\right|_{\mu = \mur}\,.
\end{eqnarray}
Notice that all kind of infinities are eliminated at once, namely
overall quadratic, overall logarithmic and overlapping logarithmic 
sub-divergences.
\section{Soft and collinear divergences}
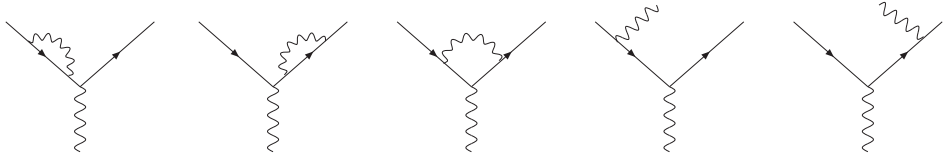
\begin{figure}
\begin{center}
\begin{picture}(450,60)(0,0)

\SetScale{0.7}
\SetOffset(55,25)
\ArrowLine(-30,35)(10,0)
\PhotonArc(-10,12)(15,-23,122){2}{6}
\ArrowLine(10,0)(50,35)
\Photon(10,0)(10,-35){3}{4}

\SetOffset(128,25)
\ArrowLine(-30,35)(10,0)
\ArrowLine(10,0)(50,35)
\PhotonArc(30,12)(15,57,202){2}{6}
\Photon(10,0)(10,-35){3}{4}

\SetOffset(203,25)
\ArrowLine(-30,35)(10,0)
\ArrowLine(10,0)(50,35)
\Photon(10,0)(10,-35){3}{4}
\PhotonArc(10,12)(15,8,178){2}{6}

\SetScale{0.7}
\SetOffset(278,25)
\ArrowLine(-30,35)(10,0)
\Photon(-18,24)(3,44){3}{4}
\ArrowLine(10,0)(50,35)
\Photon(10,0)(10,-35){3}{4}

\SetOffset(353,25)
\ArrowLine(-30,35)(10,0)
\ArrowLine(10,0)(50,35)
\Photon(40,27)(16,44){3}{4}
\Photon(10,0)(10,-35){3}{4}

\end{picture}
\caption{\label{fig:fig1} Virtual and real diagrams contributing 
to $Z \to f \bar f$.}
\end{center}
\end{figure}

FDR can also be used to regularize soft and collinear divergences. In fact, virtual and real contributions can be considered as different cuts of the same two-loop diagrams. Therefore unitarity requires, for would be massless cut lines,
\bqa
\frac{1}{q^2_j-\mu^2} \to \delta(q^2_j-\mu^2) \theta(q_j(0))\,,
\nonumber
\eqa
and the $\mu$ dependence cancels, in the {\em on-shell} limit {$\mu \to 0$}, when adding virtual and real corrections. In this section, I illustrate the simple case of the fully inclusive QED corrections to $Z \to f \bar f$ given 
in figure~\ref{fig:fig1}.
The total virtual contribution reads
\begin{eqnarray}
\Gamma_{\rm V}(Z \to f \bar f)= \Gamma_0(Z \to f \bar f)
\frac{\alpha}{\pi}\left[
-\frac{1}{2} \ln^2 \left(\frac{\mu^2}{s}\right)
-\frac{3}{2} \ln   \left(\frac{\mu^2}{s}\right)
+\frac{7}{18} \pi^2
-\frac{\pi\sqrt{3}}{2 } 
- \frac{1}{2}
\right]\,,
\end{eqnarray}
while the real part gives
\begin{eqnarray}
\label{eq:realres}
\Gamma_{\rm R}(Z \to f \bar f)= \Gamma_0(Z \to f \bar f)
\frac{\alpha}{\pi}\left[
\frac{1}{2} \ln^2 \left(\frac{\mu^2}{s}\right)
+\frac{3}{2} \ln   \left(\frac{\mu^2}{s}\right)
-\frac{7}{18} \pi^2
+\frac{\pi \sqrt{3}}{2}
+ \frac{5}{4}
\right]\,.
\end{eqnarray}
Adding the two terms gives the known result
\begin{equation}
\label{eq:totres}
\Gamma(Z \to f \bar f)
= \Gamma_0(Z \to f \bar f)\left(1+ \frac{3}{4}
\frac{\alpha}{\pi} 
\right)\,.
\end{equation}
Unlike the computation presented in~\cite{Pittau:2012zd}, 
any appearance of $\mu$ -the common vanishing mass given to all particles-  has been neglected in the numerator, keeping the $\mu$ dependence only in the propagators. This works fine in this simple QED example.
For completeness, I list, in the following, the integrals 
used in the computation 
\bqa
\label{eq:virtint}
B(s) &=&  \left. \int [d^4q] \frac{1}{(q^2-\mu^2)((q+p)^2-\mu^2)}\right|_{p^2= s} 
~=~ i \pi^2 \left[ \ln\left(-\frac{\mu^2-i\epsilon}{s}\right) +2  \right]\,, \nonumber \\
B_0 &=&  \left. \int [d^4q] \frac{1}{(q^2-\mu^2)(q^2+2(q \cdot p))}\right|_{p^2= \mu^2} 
~=~  -i \pi^2 \left( \frac{\pi}{\sqrt{3}}-2 \right)
\,, \nonumber \\
C(s)        &=& \left. \int [d^4q] 
\frac{1}{(q^2-\mu^2)(q^2+2(q \cdot p_1))(q^2-2(q \cdot p_2))}
\right|_{p_1^2= p_2^2 =\mu^2;(p_1+p_2)^2= s} \nl 
&=&  \frac{i \pi^2}{s}
\left[\frac{1}{2}\ln^2\left(-\frac{\mu^2-i\epsilon}{s}
\right)
+ \frac{\pi^2}{9} \right]
\,, \nonumber \\
C_{\rm R}  &=& \int [d^4q]  
\frac{\mu^2}{(q^2-\mu^2)(q^2+2(q \cdot p_1))(q^2-2(q \cdot p_2))}
~=~  \frac{i \pi^2}{2}\,,\nl
I_2 &=& \int_R dx dz \frac{1}{xz}
~=~ \frac{1}{2} \ln^2 \left(\frac{\mu^2}{s}\right) -\frac{7}{18}\pi^2
\,,  \nonumber \\
I_3 &=& \int_R dx dz \frac{1}{x} 
~=~ - \ln \left(\frac{\mu^2}{s}\right) -1 -\frac{\pi}{\sqrt{3}}\,, \nonumber \\
I_4 &=& \int_R dx dz \frac{x}{z} = \frac{I_3}{2} -\frac{1}{4}\,,
\eqa
where R is the full available massive three-body massive phase-space 
\begin{equation}
\int d \Phi_3 = \frac{\pi^2}{4s} \int ds_{12} ds_{23}\,,
\end{equation}
parametrized in terms of the two invariants
\begin{equation}
x= \frac{s_{12}-\mu^2}{s}\,,~~z= \frac{s_{23}-\mu^2}{s}\,.
\end{equation}
Note that, since
\bqa
{\rm Re}\left[
C(s)\frac{s}{i \pi^2}
\right] = I_2\,,
\eqa
the infrared/collinear double log is fully matched between virtual and real contributions. Furthermore -as in DR- a $\ln \mu$ of UV origin compensates a collinear log in $B_0$, leaving a finite piece.
\section{Physical interpretation and tests of FDR}
\begin{figure}
\begin{center}
\begin{picture}(300,170)(0,0)

\SetOffset(80,87)
\BCirc(-30,12){20}
\Line(-30,32)(-30,12)
\Line(-30,12)(-15,-1)
\Line(-30,12)(-45,-1)
\DashLine(-50,10)(-75,0){2}
\Photon(-30,-8)(-30,-33){2}{6}
\Gluon(-20,29)(5,48){2}{4}
\Gluon(-65,48)(-40,29){2}{4}
\Line(-30,22)(-13,14)
\Line(-8,12)(12,3)
\Gluon(-50,40)(-30,56){2}{4}
\Text(-23,-45)[tr]{(a)}

\SetOffset(220,139)
\SetWidth{1.55}
\BCirc(-30,12){20}
\Line(-30,32)(-30,12)
\Line(-30,12)(-15,-1)
\Line(-30,12)(-45,-1)
\Text(-23,-20)[tr]{(b)}

\SetOffset(220,35)
\SetWidth{0.5}
\CArc(-30,12)(20,-40,220)
\Line(-30,32)(-30,12)
\Line(-30,12)(-15,-1)
\Line(-30,12)(-45,-1)
\DashLine(-50,10)(-75,0){2}
\Gluon(-20,29)(5,48){2}{4}
\Gluon(-65,48)(-40,29){2}{4}
\Line(-30,22)(-13,14)
\Line(-8,12)(12,3)
\Gluon(-50,40)(-30,56){2}{4}
\Text(-23,-25)[tr]{(c)}

\SetWidth{1.55}
\Line(-30,5)(-15,-8)
\Line(-30,5)(-45,-8)
\CArc(-30,5)(20,220,320)
\end{picture}
\caption{\label{fig:vacuum} Generic diagram contributing to a process (a). Unphysical {\em vacuum diagrams} generated when all integration momenta are large (b)
and when one sub-loop integration momentum goes to infinity (c).}
\end{center}
\end{figure}
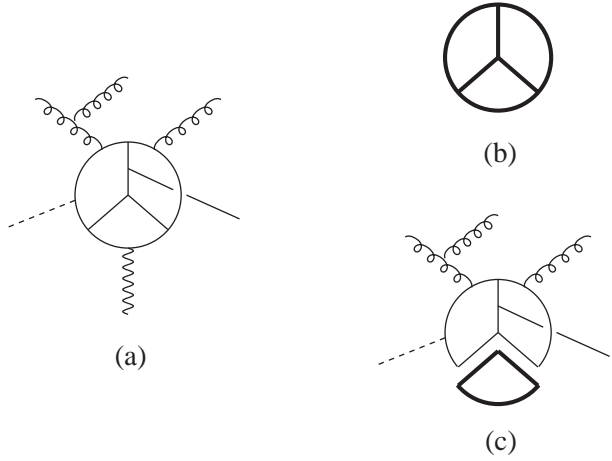

In the case of simple scalar theories, such as $\lambda \Phi^3$ and $\lambda \Phi^4$, it can be shown~\cite{Pittau:2012zd} that some of the divergent contributions, discarded in the definition of FDR integral, can be reabsorbed, at one-loop, into an unphysical vacuum expectation value of the field.
 In more complicated cases, one simply subtracts such infinities, considering that they represent an unphysical contribution to the scattering process generated when the integration momenta get large, as illustrated in figure \ref{fig:vacuum}.
The key point of this reasoning is that the part of the integration which is left, after subtraction, possesses all the relevant physical information and, in case of gauge theories, respects gauge invariance.

The latter property can be tested in realistic cases. For example, the calculation of the gluon self-energy in figure~\ref{fig:gself} (with $n_f=0$) in a general gauge with gluon propagator
\bqa
 \Delta^{\mu\nu}(p) = -i \frac{1}{p^2}
\left(
g_{\mu\nu}+(\xi-1)\frac{p^\mu p^\nu}{p^2}
\right)\,,
\eqa
gives
\bqa
\left.\Pi(p^2)\right|_{FDR} 
 = N_{col}\left(\frac{\alpha_s}{4 \pi} \right)
p^2 \left[\left(-\frac{13}{6}+\frac{\xi}{2}\right)\ln\left(-\frac{p^2}{\mur}\right) 
\left(\frac{85}{36}+\frac{\xi}{2}+\frac{\xi^2}{4}  \right)
             \right]\,,
\eqa
which is the same result one obtains in Dimensional Reduction in the $\overline{\rm MS}$ scheme. The result in Conventional Dimensional Regularization is the same, but with $\frac{85}{36}$ replaced by $\frac{97}{36}$. Known transition 
rules~\cite{Kunszt:1993sd,Pittau:2011qp} can be applied in case one needs to recover the latter scheme.  
\begin{figure}
\begin{center}
\begin{picture}(450,25)(0,0)
\SetOffset(120,0)
\Text(-38,12)[r]{$\alpha$}
\Text(-20,20)[b]{$p$}
\LongArrow(-30,18)(-10,18)
\Gluon(-35,12)(-5,12){2}{5}
\GCirc(5,12){10}{0.7}
\Gluon(15,12)(45,12){2}{5}
\Text(48,12)[l]{$\beta$}
\Text(62,12)[l]{
$= i \left(g_{\alpha \beta}-p_\alpha p_\beta/p^2\right) 
  \left. {\Pi(p^2)}\right|_{FDR}$}
\end{picture}
\caption{\label{fig:gself} Gluon self-energy in QCD.}
\end{center}
\end{figure}

As a second example, I take the loop-induced decay amplitude 
of a Higgs into two photons, $H \to \gamma(k_1^{\mu})\,\gamma(k_2^{\nu})$.
A recent FDR calculation in an arbitrary $R_\xi$ gauge~\cite{Donati:2013iya} shows full agreement with the well known result
\bqa
\label{eq_amp_form_factors}
	\amp^{\mu\nu}(\beta,\eta) &=& 	
	 	\Big( 
			\widetilde{\amp}_W(\beta)
			+\sum_f N_c Q^2_f \,\widetilde{\amp}_f(\eta)
		\Big)
		\;{T}^{\mu\nu}\,,\nl
{T}^{\mu\nu} &=& k_1^{\nu} k_2^{\mu} -(k_1\cdot k_2) \; g^{\mu\nu}\,,\nl
	\widetilde{\amp}_W(\beta) &=&
		\frac{i\,e^3}{(4\pi)^2s_W M_W}\;
		\Big[\, 
			2 + 3\beta + 3 \beta (2-\beta) f(\beta)
		\,\Big]\,, \nl
	\widetilde{\amp}_f(\eta) &=&
		\frac{-i\,e^3}{(4\pi)^2s_W M_W}\;
		2\eta\,\Big[\, 
			1 + (1-\eta) f(\eta)
		\,\Big]\,,
\eqa
with
\bqa
	\beta = \frac{4\,M_W^2}{M_H^2}\,, \qquad 
	\eta  = \frac{4\,m_f^2}{M_H^2}\,, \qquad
	f(x) = 
	-\frac{1}{4}\ln^2
		\Big( 
		\frac{1+\sqrt{1-x+i\varepsilon}}{-1+\sqrt{1-x+i\varepsilon}}
		\Big)\,.
\eqa
\section{Conclusions}
Taking the final step of {\em defining} the loop integrals in such a way that infinities simply do not occur looks promising. Such an approach is allowed as long as the definition respects shift and gauge invariance. The FDR integral obeys such properties and it is therefore a very good candidate.

One is then led to consider the difference between renormalizable and non-renormalizable theories.
When computed at $\ell$-loop in FDR, both theories give, before renormalization, Green's functions of the kind 
\bqa
\label{eq:eq23}
G^{\rm FDR}_{\ell-loop}(\mur) = \sum_{i=0}^{\ell} a_i \log^i(\mur)
+ {\rm R}(\{p,M\})\,, 
\eqa
where ${\rm R}(\{p,M\})$ depends upon the kinematical variables of the process.
When fixing the bare parameters of the Lagrangian in terms of observables, all
universal $\log^i(\mur)$ terms disappear. While no additional logarithms of the unphysical scale $\mur$ remain in the renormalizable case, no guarantee exists of their disappearance in non-renormalizable theories.
However, even in this case, one can in principle perform {\em just one} additional measurement to fix $\mur$, and obtain -at least- an effective theory valid at energy scales around the fitted value of $\mur$. Of course, nothing but the comparison with experiment can tell whether the theory is a viable one. But the problem is moved, in this way, from the occurrence of infinities to the consistency of the theory at hand. 

\bibliography{pittau}{}

\providecommand{\href}[2]{#2}\begingroup\raggedright\begin{thebibliography}{10}

\bibitem{Ossola:2006us}
G.~Ossola, C.~G. Papadopoulos, and R.~Pittau, {\it {Reducing full one-loop
  amplitudes to scalar integrals at the integrand level}},  {\em Nucl.Phys.}
  {\bf B763} (2007) 147--169,
  [\href{http://xxx.lanl.gov/abs/hep-ph/0609007}{{\tt hep-ph/0609007}}].

\bibitem{Berger:2008sj}
C.~Berger, Z.~Bern, L.~Dixon, F.~Febres~Cordero, D.~Forde, et~al., {\it {An
  Automated Implementation of On-Shell Methods for One-Loop Amplitudes}},  {\em
  Phys.Rev.} {\bf D78} (2008) 036003,
  [\href{http://xxx.lanl.gov/abs/0803.4180}{{\tt arXiv:0803.4180}}].

\bibitem{Giele:2008ve}
W.~T. Giele, Z.~Kunszt, and K.~Melnikov, {\it {Full one-loop amplitudes from
  tree amplitudes}},  {\em JHEP} {\bf 0804} (2008) 049,
  [\href{http://xxx.lanl.gov/abs/0801.2237}{{\tt arXiv:0801.2237}}].

\bibitem{Ellis:2011cr}
R.~K. Ellis, Z.~Kunszt, K.~Melnikov, and G.~Zanderighi, {\it {One-loop
  calculations in quantum field theory: from Feynman diagrams to unitarity
  cuts}},  {\em Phys.Rept.} {\bf 518} (2012) 141--250,
  [\href{http://xxx.lanl.gov/abs/1105.4319}{{\tt arXiv:1105.4319}}].

\bibitem{Mastrolia:2012an}
P.~Mastrolia, E.~Mirabella, G.~Ossola, and T.~Peraro, {\it {Scattering
  Amplitudes from Multivariate Polynomial Division}},  {\em Phys.Lett.} {\bf
  B718} (2012) 173--177, [\href{http://xxx.lanl.gov/abs/1205.7087}{{\tt
  arXiv:1205.7087}}].

\bibitem{Mastrolia:2011pr}
P.~Mastrolia and G.~Ossola, {\it {On the Integrand-Reduction Method for
  Two-Loop Scattering Amplitudes}},  {\em JHEP} {\bf 1111} (2011) 014,
  [\href{http://xxx.lanl.gov/abs/1107.6041}{{\tt arXiv:1107.6041}}].

\bibitem{Badger:2012dv}
S.~Badger, H.~Frellesvig, and Y.~Zhang, {\it {An Integrand Reconstruction
  Method for Three-Loop Amplitudes}},  {\em JHEP} {\bf 1208} (2012) 065,
  [\href{http://xxx.lanl.gov/abs/1207.2976}{{\tt arXiv:1207.2976}}].

\bibitem{Johansson:2012zv}
H.~Johansson, D.~A. Kosower, and K.~J. Larsen, {\it {Two-Loop Maximal Unitarity
  with External Masses}},  {\em Phys.Rev.} {\bf D87} (2013) 025030,
  [\href{http://xxx.lanl.gov/abs/1208.1754}{{\tt arXiv:1208.1754}}].

\bibitem{Kleiss:2012yv}
R.~H. Kleiss, I.~Malamos, C.~G. Papadopoulos, and R.~Verheyen, {\it {Counting
  to One: Reducibility of One- and Two-Loop Amplitudes at the Integrand
  Level}},  {\em JHEP} {\bf 1212} (2012) 038,
  [\href{http://xxx.lanl.gov/abs/1206.4180}{{\tt arXiv:1206.4180}}].

\bibitem{'tHooft:1972fi}
G.~'t~Hooft and M.~Veltman, {\it {Regularization and Renormalization of Gauge
  Fields}},  {\em Nucl.Phys.} {\bf B44} (1972) 189--213.

\bibitem{Freedman:1991tk}
D.~Z. Freedman, K.~Johnson, and J.~I. Latorre, {\it {Differential
  regularization and renormalization: A New method of calculation in quantum
  field theory}},  {\em Nucl.Phys.} {\bf B371} (1992) 353--414.

\bibitem{delAguila:1997kw}
F.~del Aguila, A.~Culatti, R.~Munoz-Tapia, and M.~Perez-Victoria, {\it
  {Constraining differential renormalization in Abelian gauge theories}},  {\em
  Phys.Lett.} {\bf B419} (1998) 263--271,
  [\href{http://xxx.lanl.gov/abs/hep-th/9709067}{{\tt hep-th/9709067}}].

\bibitem{delAguila:1998nd}
F.~del Aguila, A.~Culatti, R.~Munoz~Tapia, and M.~Perez-Victoria, {\it
  {Techniques for one loop calculations in constrained differential
  renormalization}},  {\em Nucl.Phys.} {\bf B537} (1999) 561--585,
  [\href{http://xxx.lanl.gov/abs/hep-ph/9806451}{{\tt hep-ph/9806451}}].

\bibitem{Battistel:1998sz}
O.~Battistel, A.~Mota, and M.~Nemes, {\it {Consistency conditions for 4-D
  regularizations}},  {\em Mod.Phys.Lett.} {\bf A13} (1998) 1597--1610.

\bibitem{Cherchiglia:2010yd}
A.~Cherchiglia, M.~Sampaio, and M.~Nemes, {\it {Systematic Implementation of
  Implicit Regularization for Multi-Loop Feynman Diagrams}},  {\em
  Int.J.Mod.Phys.} {\bf A26} (2011) 2591--2635,
  [\href{http://xxx.lanl.gov/abs/1008.1377}{{\tt arXiv:1008.1377}}].

\bibitem{Cynolter:2010ei}
G.~Cynolter and E.~Lendvai, {\it {Symmetry Preserving Regularization with A
  Cutoff}},  {\em Central Eur.J.Phys.} {\bf 9} (2011) 1237--1247,
  [\href{http://xxx.lanl.gov/abs/1002.4490}{{\tt arXiv:1002.4490}}].

\bibitem{Wu:2003dd}
Y.-L. Wu, {\it {Symmetry preserving loop regularization and renormalization of
  QFTs}},  {\em Mod.Phys.Lett.} {\bf A19} (2004) 2191--2204,
  [\href{http://xxx.lanl.gov/abs/hep-th/0311082}{{\tt hep-th/0311082}}].

\bibitem{Pittau:2012zd}
R.~Pittau, {\it {A four-dimensional approach to quantum field theories}},  {\em
  JHEP} {\bf 1211} (2012) 151, [\href{http://xxx.lanl.gov/abs/1208.5457}{{\tt
  arXiv:1208.5457}}].

\bibitem{Donati:2013iya}
A.~Donati and R.~Pittau, {\it {Gauge invariance at work in FDR: $H \to \gamma
  \gamma$}},  \href{http://xxx.lanl.gov/abs/1302.5668}{{\tt arXiv:1302.5668}}.

\bibitem{Jegerlehner:2000dz}
F.~Jegerlehner, {\it {Facts of life with gamma(5)}},  {\em Eur.Phys.J.} {\bf
  C18} (2001) 673--679, [\href{http://xxx.lanl.gov/abs/hep-th/0005255}{{\tt
  hep-th/0005255}}].

\bibitem{Kunszt:1993sd}
Z.~Kunszt, A.~Signer, and Z.~Trocsanyi, {\it {One loop helicity amplitudes for
  all $2 \to 2$ processes in QCD and N=1 supersymmetric Yang-Mills theory}},
  {\em Nucl.Phys.} {\bf B411} (1994) 397--442,
  [\href{http://xxx.lanl.gov/abs/hep-ph/9305239}{{\tt hep-ph/9305239}}].

\bibitem{Pittau:2011qp}
R.~Pittau, {\it {Primary Feynman rules to calculate the epsilon-dimensional
  integrand of any 1-loop amplitude}},  {\em JHEP} {\bf 1202} (2012) 029,
  [\href{http://xxx.lanl.gov/abs/1111.4965}{{\tt arXiv:1111.4965}}].

\end{thebibliography}\endgroup
\bibliographystyle{JHEP}
\end{document}